\input{aipcheck}

\documentclass[
  ,draft            
  ]
  {aipproc}

\layoutstyle{6x9}

\begin{document}

\title{Theoretical inputs and errors in the new hadronic currents in TAUOLA}


\author{P. Roig}{
  address={Grup de F\'{\i}sica Te\`orica, Institut de F\'{\i}sica d'Altes Energies, 
Universitat Aut\`onoma de Barcelona, E-08193 Bellaterra, Barcelona, Spain} \footnote{Speaker.}
}

\author{I. M. Nugent}{
  address={RWTH Aachen University, III. Physikalisches Institut B, Aachen, Germany}
}

\author{T. Przedzinski}{
  address={The Faculty of Physics, Astronomy and Applied Computer Science,
Jagellonian University, Reymonta 4, 30-059 Cracow, Poland},altaddress={Institute of Nuclear Physics, PAN, Cracow, ul. Radzikowskiego 152, Poland}
}

\author{O. Shekhovtsova}{
  address={IFIC, Universitat de Val\`encia-CSIC,  Apt. Correus 22085, E-46071, Val\`encia, Spain}
}

\author{Z. Was}{
  address={CERN PH-TH, CH-1211 Geneva 23, Switzerland}
  ,altaddress={Institute of Nuclear Physics, PAN, Cracow, ul. Radzikowskiego 152, Poland}
}

\begin{abstract}
The new hadronic currents implemented in the TAUOLA library are obtained in the unified and consistent framework of Resonance Chiral Theory: a Lagrangian approach in which 
the resonances exchanged in the hadronic tau decays are active degrees of freedom included in a way that reproduces the low-energy results of Chiral Perturbation Theory. 
The short-distance QCD constraints on the imaginary part of the spin-one correlators yield relations among the couplings that render the theory predictive.\\
In this communication, the obtaining of the two- and three-meson form factors is sketched. One of the criticisms to our framework is that the error may be as large as 1/3, 
since it is a realization of the large-$N_C$ limit of QCD in a meson theory. A number of arguments are given which disfavor that claim pointing to smaller errors, which 
would explain the phenomenological success of our description in these decays. Finally, other minor sources of error and current improvements of the code are discussed.
\end{abstract}

\maketitle

\section{Introduction}
New hadronic form factors have been included in TAUOLA \cite{Shekhovtsova:2012ra}, the standard Monte Carlo generator for tau lepton decays \cite{TAUOLA}. In this contribution 
we concentrate on the theoretical inputs and the associated errors. Practical aspects that may be interesting for the user are the topic of O.~Shekhovtsova's communication 
\cite{Olga}. This project is essential to meet the experimental requirements, as discussed in Ref.~\cite{Actis:2010gg}.
The definition we use for the hadronic form factors, $\mathcal{H}_\mu$, in the decay to a hadronic state $H$, $\tau^-(P)\to H \nu_\tau(N)$, is
\begin{equation}
 \mathcal{M}_\mu\,=\,\frac{G_F}{\sqrt{2}}\bar{u}(N)\gamma^\mu(1-\gamma_5)u(P)\mathcal{H}_\mu\,.
\end{equation}
We will consider the two- and three-meson hadronic states in turn.
\section{Form factors in two-meson $\tau$ decays}
For $\tau^-\to [h_1(p_1) h_2(p_2)]^- \nu_\tau$ decays, the hadronic current reads
\begin{equation}
\mathcal{H}^\mu  = N^{h_1h_2} \left[ \left(p_1 - p_2-\frac{\Delta_{12}}{s}(p_1+p_2)\right)^\mu F^{V}(s) + \frac{\Delta_{12}}{s}(p_1 + p_2)^\mu F^{P}(s) \right],
\end{equation}
where $s = (p_1 +p_2)^2$ and $\Delta_{12}=m_1^2-m_2^2$, with $N^{\pi^-\pi^0} = 1$. The other final states are given by $SU(3)$ symmetry.
The dominant contribution to these decays is given by the vector form factor, $F^{V}(s)$, which is obtained using Resonance Chiral Theory \cite{RChT} ($R\chi T$). It 
reproduces the low-energy results of Chiral Perturbation Theory \cite{ChPT} ($\chi PT$) at NLO in the chiral expansion and includes the light-flavored resonances as active 
degrees of freedom in the theory without any ad-hoc dynamical assumption. $R\chi T$ is a realization of the large-$N_C$ limit of QCD \cite{LargeNc} in a theory for the 
lightest mesons and resonances. When the asymptotic vanishing of the form factor is imposed on the result, $F^{V}(s)=1+F_V G_V/F^2\cdot s/(M_V^2-s)$, it yields 
$F^{V}(s)=M_V^2/(M_V^2-s)\equiv F^{VMD}(s)$, the vector meson dominance result.\\
FSI among the two pseudoscalar mesons are encoded in the $\chi PT$ loop function, $A_{PQ}(s)$, whose imaginary part enters the vector meson off-shell width \cite{GomezDumm:2000fz} 
included via $F^{VMD}(s)=M_V^2/(M_V^2-s-iM_V\Gamma_V(s))$. There are several ways to resum them, such as the exponentiation of the real part of the Omn\`es \cite{Omnes:1958hv} 
function \cite{Guerrero:1997ku} or the use of dispersion relations \cite{Pich:2001pj, Boito:2008fq}. The approach proposed in Ref.~\cite{Roig:2011iv} and used in 
Ref.~\cite{Shekhovtsova:2012ra}, takes Ref.~\cite{Guerrero:1997ku} as a starting point and includes phenomenologically the effect of the excited $\rho$-like resonances when 
they are seen in the data, as in the two-pion mode \cite{Belle}. The general structure for the single-resonance contribution is
\begin{eqnarray}\label{FF_2scal}
F^V_{PQ}(s) =F^{VMD}(s)\,\,\mathrm{exp}\left[\sum_{P,Q}N_{loop}^{PQ}{\frac{-s}{96\pi^2 F^2} Re A_{PQ}(s)}\right]\,,
\end{eqnarray}
where $\sum_{P,Q}$ extends over the light pseudoscalars with suitable quantum numbers to contribute in the loop and $N_{loop}^{PQ}$ is dictated by $SU(3)$ symmetry, with 
$N_{loop}^{\pi^-\pi^0}=1$.\\
The scalar form factor is important to describe correctly the low-energy region of the data \cite{Epifanov:2007rf} for the $\tau^-\to (K\pi)^-\nu_\tau$ decay \cite{JPP}. In 
the 2012 release of our code distribution for \cite{Shekhovtsova:2012ra} it has been included following the coupled channel analysis of Refs.~\cite{JOP} for the strangeness 
changing scalar form factors. The strangeness conserving scalar form factors \cite{Guo:2012yt} may be important in the modes including an $\eta$-meson.
\section{Form factors in three-meson $\tau$ decays}
In the $\tau^-\to [h_1(p_1) h_2(p_2) h_3(p_3)]^-\nu_\tau$ decays, the hadronic current reads
\begin{eqnarray}
 \mathcal{H}^\mu & = & N^{h_1h_2h_3} \left\lbrace\left(g^\mu_{\;\nu}-\frac{q^\mu q_\nu}{q^2} \right)\left[ c_1 (p_2-p_3)^\nu F_1  + c_2 (p_3-p_1)^\nu F_2  + 
c_3  (p_1-p_2)^\nu F_3 \right] \right. \nonumber\\
& & \left. + c_4  q^\mu F_4  -{ i \over 4 \pi^2 F^2} c_5 \epsilon^\mu_{.\ \nu\rho\sigma} p_1^\nu p_2^\rho p_3^\sigma F_5 \right\rbrace,
\label{fiveF}
\end{eqnarray}
where $q^\mu=(p_1+p_2+p_3)^\mu$, $F$ is the pion decay constant and $N=(V_{ij}^{CKM})/F$. Only three among the $F_1$, $F_2$ and $F_3$ form factors are independent. The 
choice for each mode, with the $c_i$ factors, can be found in Ref.~\cite{Shekhovtsova:2012ra}.\\
In this reference, the $3\pi$ and $KK\pi$ modes were included within the framework of $R\chi T$ following Refs.~\cite{Dumm:2009va} (where $\Gamma_{a_1}(q^2)$ can be found) 
and \cite{Dumm:2009kj}, respectively (the mode $\eta\pi^-\pi^0$ \cite{Dumm:2012vb} will be available soon). In these decays, vertices with more than one resonance are 
considered, including the $VVP$ vertices \cite{RuizFemenia:2003hm} ($P$ stands for one of the lightest pseudoscalar mesons) and the $VAP$ vertices \cite{GomezDumm:2003ku}. 
Some of the abundant new couplings can be related upon imposing that the imaginary part of the $VV$ and $AA$ correlators go to zero asymptotically in the large-$N_C$ limit 
\cite{Floratos:1978jb}. It is noteworthy that the relations found are in agreement for all three cases, and with those obtained in the one-meson radiative tau decays \cite{Guo:2010dv}.
\section{Discussion on the errors: As large as 1/3?}
The smallness of a expansion parameter for applying perturbation theory is given by the size of the coefficients of the expansion. Usually, in perturbation theory, i.e. QED 
or $\chi PT$, it is easy to compare the LO and NLO expressions to determine the expansion parameter and the smallness of the related coefficients warranties a good 
convergence of the lowest order computations to the true result. It is not possible to resum the infinite number of diagrams that appear in the $N_C\to\infty$ limit of QCD to 
be able to judge if the expansion parameter is small enough to rely on this approach. One can only derive that at NLO the non-planar diagrams are suppressed as $1/N_C^2$ and 
the diagrams with internal quark loops as $1/N_C$. Moreover, a factor of $n_f$ could enhance the latter diagrams, but the fact that there is negligible mixing between the 
$q\bar{q}$ and the $q\bar{q}q\bar{q}$ states hints that these kind of diagrams are heavily suppressed by their coefficients. Therefore, $1/N_C^2$ would be a better estimate 
of the effective expansion parameter, which agrees with the phenomenological success of its predictions on the hadron side. In particular, we note the good convergence of the 
predictions of the $\chi PT$ couplings working at $NLO$ in the $1/N_C$ expansion within $R\chi T$ \cite{ChPTLECsfromNLORCht} and the successful description of the hadronic 
decays of the tau lepton as indications that the expansion parameter is indeed smaller than $1/3$. Noticeably, the actual expansion parameter can be computed for $R\chi T$ in 
the study of the vector form factor of the pion at $NLO$ in the $1/N_C$ expansion \cite{VFFpiNLO}, yielding
\begin{equation}
 \frac{n_f}{2}\frac{2G_V^2}{F^2}\frac{M_V^2}{96\pi F^2}\,,
\end{equation}
which, at lowest order, is the ratio of the vector width and mass, $\sim0.2$, agreeing with the previous discussion. Moreover, we should emphasize that our approach goes 
beyond the $N_C\to\infty$ limit. We supplement the lowest order in the 1/$N_C$ expansion for the theory in terms of mesons by the leading higher-order correction, namely by 
including the resonance (off-shell) widths for the wide states $\rho$, $K^\star$ and $a_1$. This seems to point to smaller errors ($<10\%$) than those characteristic of the 
$LO$ contribution in the 1/$N_C$ expansion and may be able to explain, altogether, fine agreement with the data \footnote{It remains exciting to confront this well founded 
conjecture with experimental data in new applications.}.
\subsection{Other sources of error}
The procedure outlined in eq.(\ref{FF_2scal}) \cite{Guerrero:1997ku}, violates the unitarity and analiticity constraints at NNLO in $\chi PT$. However, it is possible to 
devise a strategy which overcomes this problem, as put forward in Ref.~\cite{Boito:2008fq}. Within it, both the real and imaginary parts of the $\chi PT$ loop functions are 
kept in the denominator of the form factor and the tangent of the relevant phaseshift, $\delta_I^J(s)$, is defined as the ratio of its imaginary and real parts and it enters 
a three-subtracted dispersion relation in order to provide the final form factor. For the $K\pi$ \cite{Boito:2008fq, JPP, Boito:2010me} and $\pi\pi$ modes \cite{Pich:2001pj, 
Roig:2011iv, Roig} the numerical differences are smaller than the experimental errors. For the decay modes where the elastic approximation is not good (such as 
K$\eta^{(\prime)}$), eq.(\ref{FF_2scal}) can be an approximation which induces a small error while a coupled channel analysis is developed. For the $K\pi$, $\pi\pi$ and 
$\pi\eta^{(\prime)}$ - which proceeds dominantly through the $\pi-\eta-\eta^{(\prime)}$ mixing \cite{Pich:1987qq}- both approaches could be employed and the error induced by 
using eq.(\ref{FF_2scal}), coded in the TAUOLA 2011 version \cite{Shekhovtsova:2012ra}, is at the percent level.\\
BaBar data for the $\tau^-\to\pi^-\pi^-\pi^+\nu_\tau$ decay \cite{Nugent:2009zz} has confirmed the nice description of Ref.~\cite{Dumm:2009va} for $d\Gamma/dq^2$ improving the 
performance of old TAUOLA hadronic currents (it happens similarly in the two-meson modes) \cite{TAUOLA2} \footnote{MC-Tester \cite{MC-Tester} is a useful tool for this kind of 
comparisons.}. However, a deviation in the low-energy region of the differential decay width as a function of the $\pi^+\pi^-$ invariant mass remains. It is possibly due to 
the absence of the $\sigma$ meson in the theoretical description. Its thorough incorporation requires the inclusion of inhomogeneities as angular averages of the form factors 
\cite{inhomogeneities} but this is very time consuming and not practical for the Monte Carlo. Instead, we include \cite{TAUOLA2} an educated parametrization of the energy 
dependence of the $I=0,2$ contributions \cite{Isidori:2006we} and use $\delta_0(s)$ and $\delta_2(s)$ consistent with the chiral constraints at low energies 
\cite{Colangelo:2001df}. The reduction of the errors in that region will be presented in Ref.~\cite{TAUOLA2}.\\
Other sources of error are smaller, such as: the contribution of SU(3) breaking terms \cite{Moussallam:2007qc}; the inclusion of excited resonances from a Lagrangian 
\cite{SanzCillero:2002bs, Mateu:2007tr}, and, in the $\pi\pi$ mode -where data are more precise- $SU(2)$ breaking \cite{SU2pipi}; and some $NNLO$ \cite{BT} subdominant terms 
which are not reproduced \cite{Guerrero:1998hd} by Guerrero-Pich-like parametrizations \cite{Guerrero:1997ku}. Their effects are discussed in more detail in 
Ref.~\cite{Shekhovtsova:2012ra}. Some improvements, along the lines discussed above will be included in the next release of our currents, see Ref.~\cite{TAUOLA2} for details.

\begin{theacknowledgments}
It is a pleasure to thank all the organizing Committee for its labor. I acknowledge instructive conversations with J.~J.~Sanz Cillero about the expansion parameter of $R\chi T$.
\end{theacknowledgments}

\bibliographystyle{aipproc}   

\end{document}